\documentclass[conference]{IEEEtran}
\IEEEoverridecommandlockouts
\usepackage{cite}
\usepackage{amsmath,amssymb,amsfonts}
\usepackage{algorithmic}
\usepackage{graphicx}
\usepackage{textcomp}
\usepackage{blindtext}
\usepackage{xcolor}
\def\BibTeX{{\rm B\kern-.05em{\sc i\kern-.025em b}\kern-.08em
    T\kern-.1667em\lower.7ex\hbox{E}\kern-.125emX}}
\makeatletter
\def\ps@IEEEtitlepagestyle{
	\def\@oddfoot{\mycopyrightnotice}
	\def\@evenfoot{}
}
\def\mycopyrightnotice{
	{\footnotesize
		\begin{minipage}{\textwidth}
			\centering
			Copyright~\copyright~ MKT'19. Personal use of this material is permitted. Permission from IEEE/VDE must be obtained for all other uses, in any current or future media, including reprinting/republishing this material for advertising or promotional purposes, creating new collective works, for resale or redistribution to servers or lists, or reuse of any copyrighted component of this work in other works.
		\end{minipage}
	}
}

\begin{document}

\title{System-Level Simulator of LTE Sidelink C-V2X Communication for 5G}
\IEEEpeerreviewmaketitle
\author{\IEEEauthorblockN{Donglin Wang}
	\IEEEauthorblockA{\textit{University of Kaiserslautern} \\
		Kaiserslautern, Germany \\
		dwang@eit.uni-kl.de}
	\and
	\IEEEauthorblockN{Raja R.Sattiraju}
	\IEEEauthorblockA{\textit{University of Kaiserslautern} \\
		Kaiserslautern, Germany \\
		sattiraju@eit.uni-kl.de}
	\and
	\IEEEauthorblockN{Andreas Weinand}
	\IEEEauthorblockA{\textit{University of Kaiserslautern}\\
			Kaiserslautern, Germany\\
			weinand@eit.uni-kl.de}	
	
	\and
	\IEEEauthorblockN{Hans D.Schotten}
	\IEEEauthorblockA{\textit{University of Kaiserslautern} \\ 
	            	\textit{German Research Center} \\ 
					\textit{for Artificial Intelligence} \\
			Kaiserslautern, Germany \\
			Hans\_Dieter.Schotten@dfki.de}
}

\maketitle

\begin{abstract}
In recent years, Cellular-Vehicle-to-Everything (C-V2X) has been an emerging area of interest attracting both the industry and academy societies to develop, which is also a prominent emerging service for the next generation of the cellular network (5G).  In the time of the development, standardization, and further improvement of 5G, so simulations are essential to test and optimize algorithms and procedures prior to their implementation process of the equipment manufactures. And C-V2X communication is used for information exchange among the traffic participants with network-assisted which can reduce traffic accidents and improve traffic efficiency. Moreover, it is also the primary enabler for cooperative driving. But C-V2X communication has to meet different Quality of Service (QoS) requirements (e.g., ultra-high reliability (99.999$\%$) and ultra-low latency). Guaranteeing a high-level reliability is a big challenge. In order to assess system performance, accurate simulations of simple setups, as well as simulations of more complex systems via abstracted models are necessary for the C-V2X communication. For checking the performance of the C-V2X communication on a highway scenario, a system-level simulator has been implemented. And, this simulation has been carried out on  the network (system-level) context. Finally, the analysis and the simulation results for the C-V2X communication are presented, which shows that different objectives can be met via system-level simulation.
\end{abstract}

\begin{IEEEkeywords}
	5G, C-V2X communication, System-level
\end{IEEEkeywords}

\section{Introduction}
In wireless mobile communication, realistic measurement approaches are becoming costlier and more laborious in the real world. So it’s extremely inevitable and essential to simulate the communication network for figuring the mutual information interactions of all involved elements. Especially, simulations are more efficient for gaining insights into the performance of both small-and-large scale scenarios. Therefore, simulations are carried out [1]. Our system-level simulator is developed along with the standard-compliant Evolved Universal Terrestrial Radio Access (E-UTRA) / Long Term Evolution (LTE) 3rd Generation Partnership (3GPP) release-14 “Sidelink” radio technology and its integration into 5G network infrastructures [2]. Also, sidelink is known as PC5 which can facilitate the direct communication between the two ends of a C-V2X communication link. Up to 3GPP release 14, the sidelink communication is connection-less which means there is no Radio Resource Control (RRC) connection over PC5 air interface. Since network infrastructure is not involved in the user-plane data transmission with direct C-V2X communication, packet transmission latency can be efficiently reduced. In comparison, in [3], C-V2X communication transmitting data packets through the cellular network infrastructures has been considered. LTE-Uu interference facilitates the C-V2X communication and a higher end-to-end (E2E) latency compared to the direct C-V2X communication can be foreseen. Moreover, a lot of simulators have been implemented (e.g. Vienna Simulators) which focus on the LTE downlink and uplink system-level simulators [4][5].  Following the ongoing discussion on the 5G, we introduce the new 5G sidelink simulators. In this work, the performances of the direct C-V2X communication has been provided through the system-level simulator. 
At the beginning of this work, we described the communication system. In Sect. III, we gave an overview of the capabilities and the general purposes of our system-level simulator, introduce structure and describe the key features. Following that, the system-level simulator implementation details are described in Sect. IV. In V, we got these simulation results of the sidelink LTE C-V2X communications and inspected the performances of this cellular wireless communications, and comparisons have done in order to analyze QoS requirements.

\section{System description}
	\begin{figure*}[htbp]
		\centering
		\includegraphics[width=\linewidth]{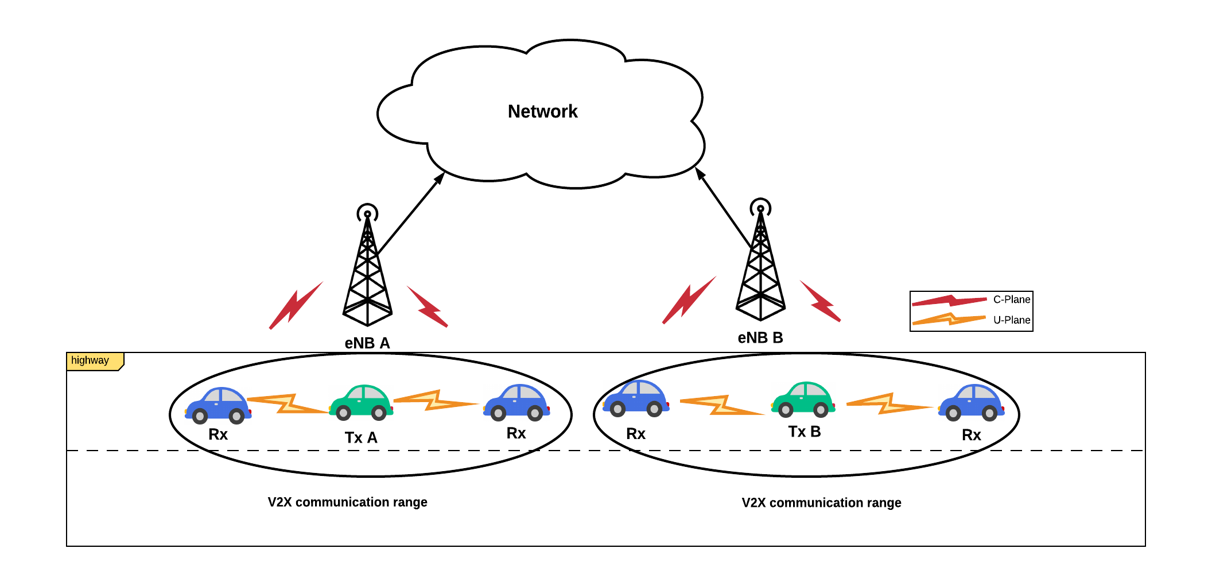}
		\caption{Direct C-V2X Communication with Network Assistance on a Highway}
		\label{fig}
	\end{figure*}
The direct C-V2X communication through sidelink is a mode of communication whereby a User Entity (UE) can directly communicate with other UEs in its proximity over the PC5 air interface proposed in 3GPP [6]. This communication is a point-to-multipoint communication where several Rxs try to receive the same data packets transmitted from the Tx. As shown in Fig.1, network-assisted direct C-V2X transmission model is implemented in this work under a highway scenario for data packets transmission. And all UEs are connected to
Base Stations (BSs) and the UE radio architecture consisted of the U-Plane and C-Plane is provided for the C-V2X communication [7]. In this work, the Tx directly transmits its data packets to the surrounding Rxs in the communication range of the Tx in the U-Plane. Therefore, without involving the C-Plane, the direct C-V2X communication is efficient from the latency aspect. And all UEs are connected to the operator network in the C-Plane where can provide network control for the direct C-V2X communication. In addition, a prerequisite is that the UE must be authorized and allowed to use the Proximity Services (ProSe) which is the logical function used for network related actions required for ProSe [8]. In 3GPP, there are two sidelink transmission modes to assign radio resources to C-V2X Txs. 
	\begin{itemize}
	\item Sidelink transmission mode 3: This transmission mode is only available when the vehicles are under cellular coverage. To assist the resource allocation procedure at the BS, UE context information (e.g., traffic pattern geometrical information) can be reported to BSs.
	\end{itemize}

	\begin{itemize}
	\item Sidelink transmission mode 4: In this mode, a Tx in C-V2X communication can autonomously select a radio resource from a resource pool which is either configured by network or pre-configured in the user device for its direct C-V2X communication over PC5 interface. In contrast to mode 3, transmission mode 4 can operate without cellular coverage. 
	\end{itemize}
	In this work, only transmission mode 3 is utilized for the direct C-V2X communication through sidelink which means all UEs are under the coverage of the cellular network and network control the resource allocation. 
\section{Sidelink System-level Simulator}

 Usually, System-level simulation is used to measure the performance of cell networking and simulating large networks comprising multiple Enhanced Node-Bs (eNBs) and UEs [3]. And the system relevant functionalities like network layout, channel model, and characters of a BS and a mobile UE are taken into account for providing a good insight into the system-level simulator. System-level simulation is generally based on multi-cell multi-user, so its channel considers large-scale fading at the same time. Thus, the UE will distinguish the distance from the BS, and consider the interference situation of different neighboring areas, because it is a multi-UE system. However, for system-level simulation of the sidelink C-V2X communication, data packets are directly transmitted from a Tx to the Rx without going through the network infrastructure. BSs are only responsible for providing the control information to UEs under coverage. And the interferences are from the UEs who are utilizing the same transmission resource and transmitting data packets simultaneously. Moreover, the platform mainly reflects the Media Access Control (MAC) layer algorithm such as resource allocation, user scheduling, and adaptive modulation and coding scheme, rather than that the physical layer processing, the performance of the physical link is taken into account by a Link-to-System (L2S) interface based on link level simulations[1]. Moreover, to precisely reflect the characteristics of a radio link (e.g., frequency fading), a L2S level mapping needs to be accurately formulated[6]. Further, mutual information based L2S is one of the commonly used methods which has been considered as preferable and applicable. Physical-layer procedures have to be abstracted by accurate but also low-complexity models. In our system-level simulation, UE capacity, scheduling, mobility management, admission control, and interference management are Investigated. In the simulation-based approach, the Key Performance Indicators (KPIs) need to be derived by running a simulator taking account of the different features in the relevant protocol layers, e.g., modulation and coding scheme (MCS), hybrid automatic repeat request (HARQ), and scheduling. In addition, Bit Error Rate (BER), Block Error Rate (BLER), Signal to Noise ratio (SNR), Signal-to-Noise-plus–Interference-Ratio (SINR), and Packet Reception Ratio (PRR) describing the system performance are all inspected and simulated in the system-level simulator [9].

\section{Procedures of system-level simulator}
In order to evaluate the sidelink C-V2X communication, the system-level simulator has been implemented to inspect on the performance of this communication. We provide a detailed system-level simulation chain and parameters in the following Fig.2. And other system detailed information and guideline are provided in [7].

	\begin{figure}[htbp]
	\centering
	\includegraphics[width=\linewidth]{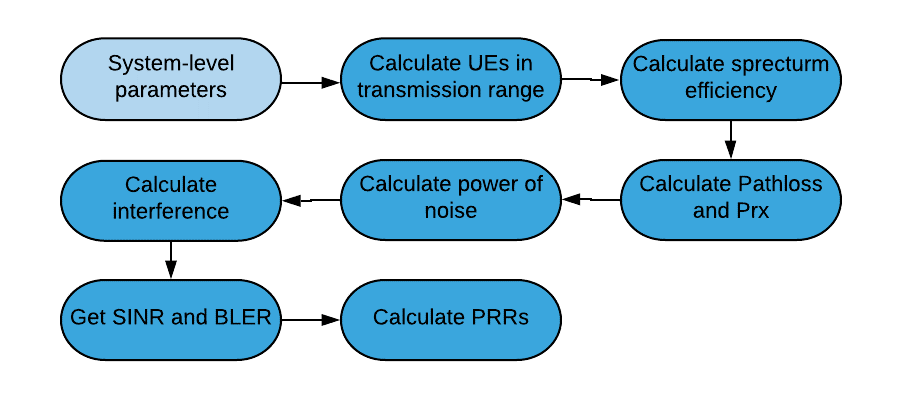}
	\caption{System-level simulator chain}
	\label{fig}
	\end{figure} 

	\subsection{Simulation Assumptions}
	There are plenty of parameters utilized in the system-level simulator as shown in Tab.I.
		\begin{itemize}
			\item Environment Model
		\end{itemize}
	BSs are deployed with an Inter-Site-Distance (ISD) of 1732 meters alongside the 3-kilometer highway scenario with 6 lanes to provide the network connections to the UEs of the C-V2X communication. And we assume the communication range of the UEs is 400 meters [7].
		\begin{itemize}
			\item Deployment Model
		\end{itemize}
	In this work, the antenna height of BS utilized is 35 meters [7]. And an isotropic antenna is installed on the top of each vehicle at a height of 1.5 meters. And 1×2 antennas configuration is exploited for the direct C-V2X communication. Also each direct C-V2X Tx has a constant transmission power of 24 dBm. Moreover, the central carrier frequency is 5.9 GHz with a transmission bandwidth of 10 MHz.
		\begin{itemize}
			\item Traffic Model
		\end{itemize}
	Here, we use a periodic package transmission of 212 Bytes with 10 Hz periodicity for each vehicle [11].
		
		\begin{itemize}
			\item Channel Model
		\end{itemize}
		The WINNER II models are applied for evaluating the performance of the sidelink C-V2X communication. The WINNER II channel models [11] are the propagation models for calculating the pathloss. In this work, the WINNER II channel models we utilized to inspect on the C-V2X communication are the following:
\begin{multline}
PL_{LOS} = 21.5log10(d) +20log10(\frac{f_c}{5.0}),   \hspace{0.5cm}   \sigma= 4,\\
if \hspace{0.2cm} 10m < d < d_{BP}, \label{eq}
\end{multline}
\begin{multline}
PL_{LOS} =  40log10(d) +10.5 - 18.5log10(h_{BS})-   \\
18.5log10(h_{MS})+1.5log10(\frac{f_c}{5.0}),   \hspace{0.5cm}   \sigma= 6,\\
if \hspace{0.2cm} d_{BP} < d < 10km, \label{eq}
\end{multline}	
\begin{multline}
PL_{NLOS} =  25.1log10(d) +55.4 - 0.13log10(h_{BS}-    \\
25)log10(\frac{d}{100})-0.9(h_{MS}-1.5)+21.3log10(\frac{f_c}{5.0}), \hspace{0.5cm}   \sigma= 8, \\
if \hspace{0.2cm} 50m< d < 5km, \label{eq}
\end{multline}
		Where $fc$ is the central frequency in 5.9 GHz and $d$ is in meters. $d_{BP}$ is the breakpoint and computed as  $d_{BP}$  =4\(h_{BS}\)\(h_{MS}\)\(fc/c\) where $fc$ is in Hz. $h_{MS}$ and $h_{BS}$ are the antenna heights of the mobile station and BS. The shadowing models for the three cases are added to the direct transmission links which are log-normal random variables with 4 dB, 6 dB, and 8 dB standard deviations respectively.
		
		\begin{table}[htbp]
		\caption{General simulation parameters of system-level simulator}
		\begin{center}
			\begin{tabular}{|l|l|}
				\hline
				Simulation parameters & Values\\ 
				\hline
				Scenario & Highway \\ 
				\hline
				Cellular layout & ISD 1732m  \\ 
				\hline
		    	Length of Highway  & 	3000 meters  \\
		    	\hline
				Number of lanes  & 	6   \\
				\hline
				Lane width & 	4 meters    \\
				\hline
				Antenna height of BS  & 	35 meters   \\
				\hline
				Antenna height of UE  & 	1.5 meters   \\
				\hline
				Channel bandwidth  & 	10 MHz   \\
				\hline
				Carrier frequency & 	5.9 GHz   \\
				\hline
				Modulation 	 & broadcast: adaptive MCS   \\
				\hline
				Data traffic  & 	Highway: 10 Hz (CAM)   \\
				\hline
				Packet size  & 	256 bytes   \\
				\hline
				Tx power & 	24 dBm   \\
				\hline
				Required communication range  & 	400 meters   \\
				\hline
				Required PRR  & 	$\geq$ 99 \%     \\
				\hline
				Vehicle density or IVD  & 	54 vehicles/cell or 100 meters   \\
				\hline
				Vehicle speed & 	100 km/h or 140 km/h   \\
				\hline
				Noise figure & 	9 dB   \\
				\hline
				Channel model & 	WINNER II channel model   \\
				\hline
				Tx antenna gain	 & 0 dB   \\
				\hline
				Rx antenna gain	 & 3 dB   \\
				\hline
			\end{tabular}
		\end{center}
	\end{table}
	\subsection{Calculations}
	All parameters utilized in the direct C-V2X communication have been provided.
		\begin{itemize}
			\item Number of UEs
		\end{itemize}
		As shown in Fig.1, we take the Tx A under the control of eNB A as an example. The amount of the Rxs in the communication range of Tx A is summed.
		\begin{itemize}
			\item Modulation and Coding Scheme
		\end{itemize}
	    First of all, the data volume value is calculated as:
			\begin{equation}
			DataVolume=P\times UE\times TF
			\end{equation}
		Where the $P$ is the packet size, $UE$ is the number of the utilized users. And $TF$ is the packet transmission frequency.
		An appropriate MCS is quite crucial for a point-to-multipoint communication and should be able to meet the system capacity requirement and provide a good robustness. The Spectral Efficiency (SE) is calculated as:
		\begin{equation}
		SE = DataVolume /BW \label{eq}
		\end{equation}
		Where $DataVolume$ is given in Eq.4. The $BW$ is the system transmission bandwidth. It is also worth noting MCSs with different SEs are applied in LTE and each MCS provides different robustness as shown in Fig.3.
		\begin{figure}[htbp]
			\centering
			\includegraphics[width=\linewidth]{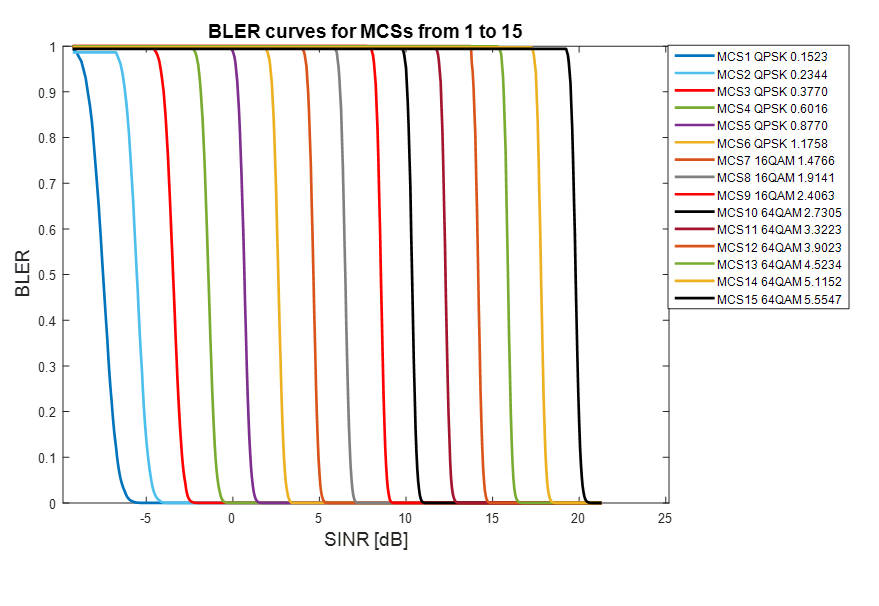}
			\caption{Mapping SINR to BLER [3]}
			\label{fig}
	    \end{figure} 
	    \begin{figure*}[htbp]
	    	\centering
	    	\includegraphics[width=\linewidth]{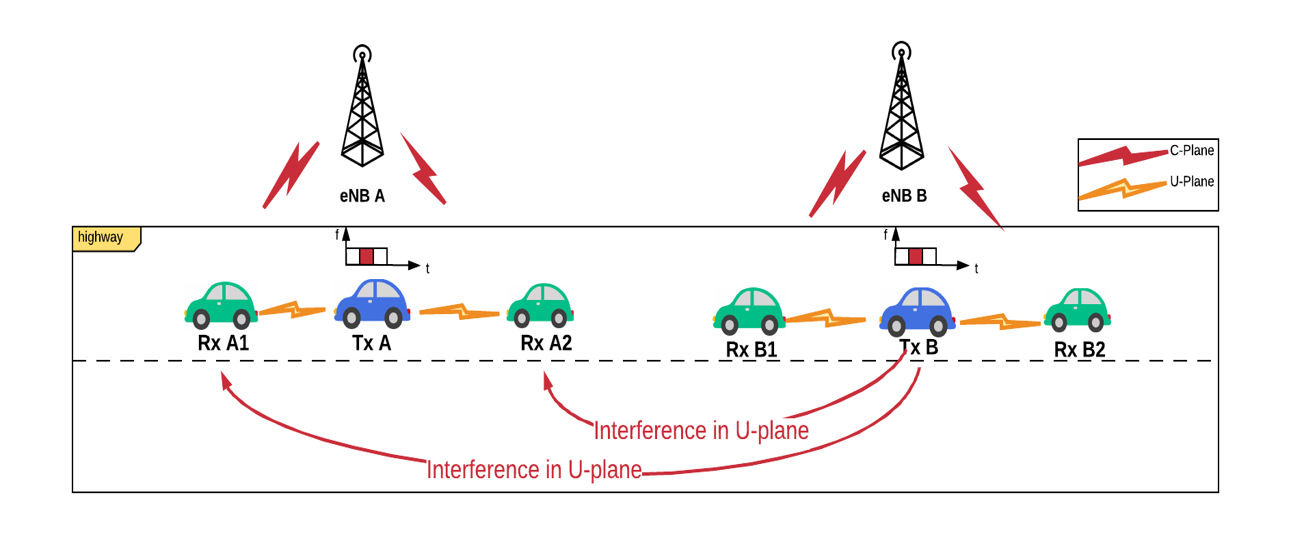}
	    	\caption{Interference of the direct C-V2X communication}
	    	\label{fig}
	    \end{figure*}
		\begin{itemize}
			\item Received Power of a Rx
		\end{itemize}
		In this work, only large-scale fading like shadow fading is taken into consideration. So the received power of the Rx in the communication range of the Tx is calculated as: 
		\begin{equation}
		P_{rx}=P_{tx} + G_{Rx_a} + G_{Tx_a} - PL
		\end{equation}	
		Where $P_{tx}$ is the transmission power (dBm), $G_{Rx_a}$ and $G_{Tx_a}$ (dBi) are the antenna gains of receiver and transmitter respectively. $PL$ is the pathloss which can be calculated from Eq.(1)(2)(3).
		\begin{itemize}
			\item Noise Figure
		\end{itemize} 
		Noise power in an Rx is usually dominated by the thermal noise [10]. As shown in Tab.1, thermal noise Power Spectral Density (PSD) is -174 dBm/Hz. The experienced noise power of the received signal can be calculated as:
		\begin{equation}
		P_{N}=Thermal_{N} + NF + BW
		\end{equation}	
		where $NF$ is the noise figure of the Rx, and $BW$ is the logarithm of the reception radio bandwidth resource.
		\begin{itemize}
			\item Interference 
		\end{itemize}	
        For sidelink C-V2X communication, data packets are directly transmitted from a Tx to the Rx without going through the network infrastructure. BSs are only responsible for providing the control information to UEs under coverage. Interference is from the Txs who are utilizing the same transmission bandwidth resource and transmitting data packets simultaneously. In Fig.4, Both Tx A and Tx B are using the same frequecy resource for data packets transmission. Rx A1 and Rx A2 are in the communication range of Tx A. Thus, interference coming from Tx B can be experienced by Rx A1 and Rx A2 which try to receive the data packets from Tx A.
		\begin{itemize}
			\item SINR and BLER
		\end{itemize}	
	    The SINR values of the Rxs are calculated as:
			\begin{equation}
			SINR= \frac {P_{R_x}} {P_{inter}+P_{noise}}
			\end{equation}
		Where $P_{Rx}$ is the receiving power of the wanted Tx, and the $P_{inter}$ and $P_{noise}$ are the powers of the interference and noise respectively. The SINR is mapped to BLER according to Fig. 3. 
		\begin{itemize}
			\item Packet Reception Ratio
		\end{itemize}
		PRR is defined as a percentage of nodes that successfully receive a packet from the tagged node among the Rxs that are within transmission range of the Tx at the moment that the packet is sent out [13]. In this work, a BLER value of 1\% is set as the threshold. Tx searches for all Rxs which have an calculated BLER less than a pre-defined threshold value. Then, we get the PRR values to represent the system performance of the direct C-V2X communication.

\section{Simulation results analysis}

In this section, the simulation results obtained from the system-level simulator are provided. PRRs are utilized to represent system performances of the sidelink C-V2X communication on the highway scenario.  

\begin{table}[htbp]
	\caption{Calculations of the simulation parameters}
	\begin{center}
		\begin{tabular}{|c|c|c|c|c|}
			\hline
			IVD    & UEs    & Data volumes Mbps   & MCS SE b/s/Hz & PRR  \%      \\ 
			\hline
			5      & 1920   & 39.3216    & 4.5234  &  80.36   \\ 
			\hline
			10     & 960    & 19.6608   & 2.4063  &  81.94  \\ 
			\hline
			20     & 480    & 9.8304    & 1.1758  &  88.77  \\
			\hline
			40     & 240    & 4.9152    & 0.6016  &  90.83  \\
			\hline 
			40     & 240    & 4.9152    & 0.6016  &  90.83  \\
			\hline
			50     & 192    & 3.9322    & 0.6016  &  90.90  \\
			\hline
			80     & 120    & 2.4576    & 0.3770  &  93.81  \\
			\hline
			100    & 96     & 1.9660    & 0.2344  & 96.40   \\
			\hline
		\end{tabular}
	\end{center}
\end{table}

 The Tab.II shows the calculated data volumes, SEs, and PRRs  of LTE sidelink C-V2X communication of different IVDs for 10 Hz transmission frequency. It's easy to find that PRR increases when the IVD is increased as well. The PRR increases from 80.36$\%$ to 96.40$\%$ with the IVD increasing from 5 meters to 100 meters. An increased IVD means fewer vehicles will be deployed on the highway scenario. Since a lower number of vehicles introduces a lower system load, as shown in Tab.II, an MCS with a better robustness can be applied. that’s why PRRs are raised with the increased IVDs when vehicles are transmitting with the same frequency resource. In Fig.5, the PRRs of the different traffic loads for the sidelink C-V2X communication are plotted where the communication range is set to be 400 meters. And the PRR values can be increased when vehicles are transmitting with a decreased traffic load. For example, when we apply an IVD of 5 meters, the PRR is increasing from 80$\%$ to 91$\%$ with decreasing traffic loads. The reason is, in Eq.4, we can easily find if we decline the transmission frequency, the data volume will be declined too. With the decreased data volume, we can apply a MCS with better robustness.
		\begin{figure}[htbp]
			\centering
			\includegraphics[width=\linewidth]{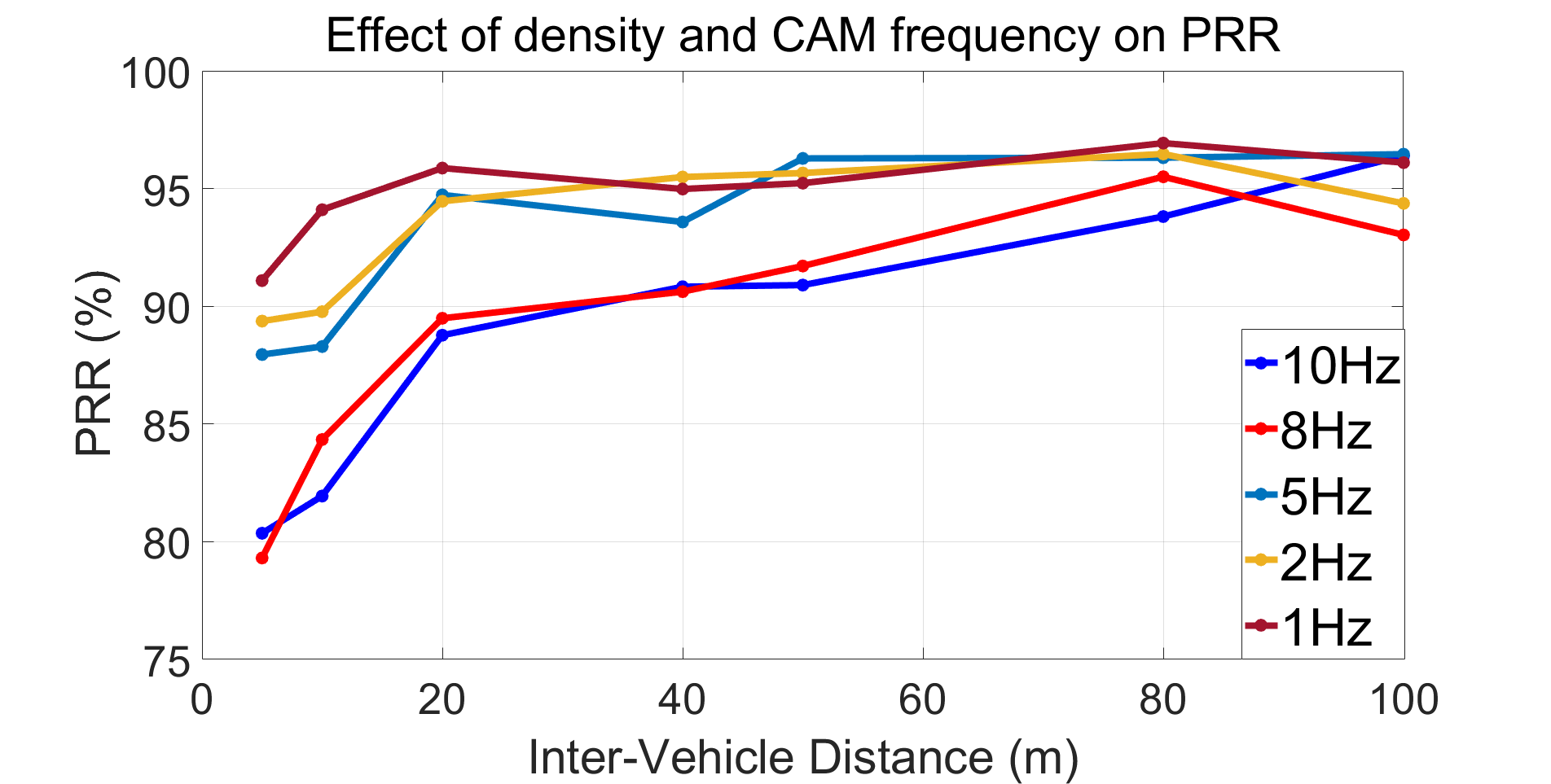}
			\caption{IVDs and Traffic loads Vs PRRs}
			\label{fig}
		\end{figure}  
\section{Conclusion}
The implementations of sidelink system-level simulator for the direct C-V2X communication on the highway scenario is clearly introduced in this work. In the system level simulator of the direct C-V2X communication on this highway, we implement the whole network and get data volumes, MCS SEs, and PRR values in order to inspect the performance of the direct C-V2X communication. And we have provided a detailed analysis of the traffic load of this communication system. In addition, we utilize different IVDs for analyzing the influence of the communication system capability. 

\section{acknowledegment}

Part of this work has been performed in the framework of the Federal Ministry of Transport and Digital Infrastructure (BMVI) project ConVeX. The authors would like to acknowledge the contributions of their colleagues, although the views expressed are those of the authors and do not necessarily represent the project.

\end{document}